\documentclass[12pt]{article}
\usepackage{amsmath,amssymb}
\usepackage[english]{babel}

\begin{document}
\vspace*{-2cm}
\hfill   \hbox{\bf SB/F/02-301}
\hrule \vskip 2cm

\begin{center}
{\Large\bf{New results concerning the $so(2,1)$ treatment for the
hypergeometric Natanzon potentials}}\\ \vspace*{0.3cm}
{\large{Sebasti\'{a}n Salam\'{o}\footnote{E- mail address:
ssalamo@fis.usb.ve}  \footnote{Invited talk given at the ``III
congreso de la Sociedad de F\'{\i}sica'', 10-14 diciembre 2001,
Caracas, Venezuela.} \footnote{accepted for publication in 
Rev. Mex. F\'{\i}sica, 2002.}}}\\
{\small{Universidad Sim\'{o}n Bol\'{\i}var, Departamento de F\'{\i}sica,}}\\
{\small{Apartado Postal 8900, Caracas, Venezuela}}
\end{center}

\begin{abstract}
The $so(2,1)$ analysis for the bound state sector of the
hypergeometric Natanzon potentials (HNP) is extended to the
scattering sector by considering the continuous series of the
$so(2,1)$ algebra. As a result a complete algebraic treatment of
the HNP by means of the $so(2,1)$ algebra is achieved.

In the bound state sector we discuss a set of satellite potentials
which arises from the action of the $so(2,1)$ generators. It is
shown that the set of new potentials are not related to the one
obtained by means of SUSYQM or of the potential algebra approach
using the $so(2,2)$ algebra.
\end{abstract}

\section{Introduction}

\indent Algebraic techniques have been developed in the last
decade to describe the bound state sector for Natanzon potentials
[1] in their two forms; the confluent and the hypergeometric ones.
The HNP was treated by means of an $so(2,2)$ algebra, the so
called potential group approach developed in [2]. More recently
the HNP was analyzed using the $so(2,1) $ algebra [3], the
confluent case also admits an a $so(2,1)$ algebra in the
description of the bound sector [4].

The scattering sector\ for Coulomb problem has also been treated
by using group theoretic methods long time ago [5]. Later on, in
[6] a technique was\ developed for systems whose Hamiltonian may
be written as a function of the Casimir invariant of an $so(2,1)$
algebra. They have developed a purely algebraic technique for the
calculation of the $S$ matrix, which they call Euclidean
connection, after noticing that scattering states are related to
the eigenstates of the $e(2)$ algebra. The same authors, [2] where
able to extend the Euclidean connection to deal with the
scattering problem for the HNP. Another point of view for deal
with systems whose Hamiltonian can be written in terms of an
$so(2,1)$ algebra was developed in [7].

\indent In this article we extend the algebraic treatment of the
bound states of the HNP, developed in [3], to the scattering
sector. Analyzing the asymptotic behavior of this particular
realization of $so(2,1)$ and using the formalism developed in [8],
we derive the $S$ matrix for the HNP. The simplicity of this
treatment is stressed out.

\indent The organization of this work is as follows: a) A brief resume of
the $so(2,1)$ analysis for the bound state sector for the HNP. An
example is developed, the P\"{o}schl-Teller potential. b) The
asymptotic algebra is shown to be an $so(2,1)$ algebra. Then the
scattering of the P\"{o}schl-Teller potential is analyzed. c) The
general case of the HNP is treated. Finally we discuss new
satellite potentials related with \ the approach given in [3]

\section{The bound state sector}

The HNP are given by [1]

\begin{eqnarray}
V(z) &=&\frac{f\,z^{2}-\left( h_{0}-h_{1}+f\right)
z+h_{0}+1}{R(z)} \label{1} \\ &&+{\left( a+\frac{a+\left(
c_{1}-c_{0}\right) \left( 2z-1\right) }{z\left( z-1\right)
}-\frac{5}{4}\frac{\Delta }{R(z)}\right) }
\frac{z^{2}\left( 1-z\right) ^{2}}{R(z)^{2}}  \notag
\end{eqnarray}

where
\begin{eqnarray}
R(z) &=&a\,z^{2}+\tau \,z+c_{0},\quad \tau =c_{1}-c_{0}-a,\ \
\label{2} \\ \Delta  &=&\tau ^{2}-4\,a\,c_{0}  \notag
\end{eqnarray}

\indent The constants $f$, $h_{0}$, $h_{1}$, $a$, $c_{0}$, $c_{1}$
are called Natanzon parameters. The function $z$ is supposed to
depend on the variable $r$ and satisfies
\begin{equation}
\frac{d\,z}{dr}=\frac{2\,z\left( z-1\right) }{\sqrt{R(z)}},
\label{3}
\end{equation}
the function $z$ is restricted to $\left[ 0,1\right] $ and the
transformation $r(z)$ is assumed to carry $r\rightarrow \infty $ to $%
z=1,r\rightarrow 0$ to $z=0$ [1].

\indent In the algebraic description of the HNP using $\ $an
$so(2,1)$ algebra [3] a two-variable realization of the algebra is
used. The Hamiltonian $H$ is related to the Casimir invariant $Q$
via $(Q-q)\Psi (r,\phi )={\mathcal{G}}(r)(E-H)\Psi (r,\phi )$,
where $q$ and $E$ are the eigenvalues of $Q$ and $H$.
${\mathcal{G}}(r)$ is a function fixed by consistency. The
eigenfunctions of the Hamiltonian have the form $\Psi (r,\phi
)=e^{im\phi }\Phi (r)$. The realization of $so(2,1)$ used is
\begin{eqnarray}
J_{0} &=&-i\frac{\partial }{\partial \phi }  \notag \\
J_{\pm } &=&e^{\pm i\phi }\left( \pm \frac{\sqrt{z}(z-1)}{z^{\prime }}\frac{%
\partial }{\partial r}-\frac{1}{2}\frac{i(z+1)}{\sqrt{z}}\frac{\partial }{%
\partial \phi }\right.   \label{4} \\
&&\left. {\pm }\frac{1}{2}(z-1)\left[ \frac{1\mp p}{\sqrt{z}}-\frac{z"\sqrt{z%
}}{z^{\prime }{}^{2}}\right] \right) ,  \notag
\end{eqnarray}
where $z^{\prime }=dz/dr$ and $p$ is a function of the Natanzon
parameters. This generators satisfy the usual commutation
relations of the $so(2,1)$
algebra: $\left[ J_{0},J_{\pm }\right] =\pm J_{\pm }$, $\left[ J_{+},J_{-}%
\right] =-2\,J_{0}$. The Casimir operator $Q$ turns out to be

\begin{eqnarray}
Q &=&\frac{z\left( z-1\right) ^{2}}{z^{\prime }{}^{2}}\frac{\partial ^{2}}{%
\partial r^{2}}+\frac{1}{4}\frac{\left( z-1\right) ^{2}}{z}\frac{\partial
^{2}}{\partial \phi ^{2}}  \notag \\
&&+\frac{1}{2}\frac{i\,p\left( z^{2}-1\right) }{z}\frac{\partial
}{\partial \phi }  \label{5} \\ &&+\frac{1}{4}\left( z-1\right)
^{2}\frac{\left[ z^{2}\left( 2z^{\prime \prime \prime }z^{\prime
}-3z"\right) -z^{\prime }{}^{4}\left( p^{2}-1\right) \right]
}{z\,z^{\prime }{}^{4}}  \notag
\end{eqnarray}

We should emphasize that the generators exhibit in (4) depend on
the parameter $p$ that in general depend on the label $\nu $, the
same one that occurs in the energy spectra.

\indent The eigenvalues of the compact generator $J_{0}$ are known
to be
\begin{equation}
m_{\nu }=\nu +\frac{1}{2}+\sqrt{q_{\nu }+\frac{1}{4}},\,\nu
=0,1,... \label{6}
\end{equation}
where $q_{\nu }$ are the eigenvalues of the Casimir operator $Q.$
The energy spectra is given by
\begin{equation}
2\nu +1=\alpha _{\nu }-\beta _{\nu }-\delta _{\nu }  \label{7}
\end{equation}
where
\begin{eqnarray}
\alpha _{\nu } &=&\sqrt{-aE_{\nu }+f+1}=p_{\nu }+m_{\nu }  \notag
\\ \beta _{\nu } &=&\sqrt{-c_{0}E_{\nu }+h_{0}+1}=p_{\nu }-m_{\nu
}  \label{8}
\\
\delta _{\nu } &=&\sqrt{-c_{1}E_{\nu }+h_{1}+1}=\sqrt{4q_{\nu
}+1}.  \notag
\end{eqnarray}

The set $\left\{ p_{\nu },q_{\nu },m_{\nu }\right\} $are called
group parameters, these label the states of the system. The
carrier space for the given representation is
\begin{eqnarray}
\Phi _{p_{\nu }q_{\nu }m_{\nu }} &=&K\ z^{\beta _{\nu
}/2}(1-z)^{(\delta _{\nu }+1)/2}{}
_{2}F_{1}(-\nu ,\alpha _{\nu }-1,1+\beta _{\nu },z),\label{9} 
\end{eqnarray}
where $K$ is a normalization constant.

To fix ideas let us consider as an example the P\"{o}schl-Teller
potential given by
\begin{eqnarray}
V_{pt}=-A(A+1){{\mathrm{sech}}}(r)^{2}+B(B-1){{\mathrm{csch}}}(r)^{2}
\label{10}
\end{eqnarray}
It is a simple task to verify that the Natanzon parameters given
by
\begin{eqnarray}
a &=&c_{0}=0,\quad c_{1}=1  \notag \\ \ h_{0}
&=&\frac{(2B+1)(2B-3)}{4}  \label{11} \\ h_{1} &=&-1,\quad
f=\frac{(2A-1)(2A+3)}{4}  \notag
\end{eqnarray}
with $z={\tanh} (r)^{2}$ reproduce (10). For the energy spectra we
found after using (7)
\begin{equation}
E_{\nu }=-(2\nu -A+B)^{2},\;\nu =0...\nu _{\max },  \label{12}
\end{equation}
where $\nu _{\max }=$intpart$(A-B)/2,$ we assume $A>B$ in order to
have bound states. The result given in (12) is obtained by a
careful study of the ambiguities occurring in (8) due of the signs
of the square roots involved, the main point is that the energy
should increase with $\nu .$ The group parameters are
\begin{eqnarray}
p_{\nu } &=&\frac{A+B}{2},\ m_{\nu }=\frac{A-B+1}{2}  \label{13}
\\ q_{\nu } &=&\frac{(2\nu +1-A+B)(2\nu -1-A+B)}{4}.  \notag
\end{eqnarray}

The generators and the Casimir operator are obtained from (4) and
(11), we obtain

\begin{eqnarray}
J_{pt\pm } &=&\left( \pm \frac{\partial }{\partial
r}-\frac{i(1+{\tanh} (r)^{2})}{2{\tanh} (r)}\frac{\partial
}{\partial \phi }+\right.   \label{14} \\ &&\left. \frac{((2p\pm
1){\tanh} (r)^{2}-2p\pm 1)}{4{\tanh} (r)}\right) \exp (\pm i\phi )
\notag
\end{eqnarray}

\begin{eqnarray}
Q_{pt} &=&\frac{1}{4}\frac{\partial ^{2}}{\partial
r^{2}}-\frac{ip_{\nu }(1-{\tanh} (r)^{4})}{2{\tanh} (r)^{2}}
+\frac{(1-{\tanh} (r)^{2})}{4{\tanh} (r)^{2}}\frac{\partial
^{2}}{\partial \phi ^{2}}\\ 
&&+\left[(1-4p_{\nu }^{2})(1+{\tanh}
(r)^{4})+(8p^{2}-6){\tanh} (r)^{2}\right]\frac{1}{16{\tanh} (r)^{2}}  \notag
\end{eqnarray}

With these results the bound sector has a complete group
description. The next step is an algebraic treatment of the
scattering sector for the example given.

\section{Scattering sector}

We have seen in the previous section that the algebra for
describing the bound state sector is an $so(2,1)$ one. For the
scattering sector we first analyze the case of the P\"{o}schl-Teller
potential. Following the ideas developed in [8], one can ask for
the asymptotic limit, $r\rightarrow \infty ,$ of the bound state
algebra. One can guess that the limiting algebra could be suitable
to describe the scattering sector, we are going to see that indeed
this is the case. We define the asymptotic algebra as the limits
of the one given in (4,5), after a straightforward calculation we
obtain
\begin{eqnarray}
J_{\pm }^{\infty } &=&\exp (\pm i\phi )\left[ -i\frac{\partial
}{\partial
\phi }\mp \frac{\sqrt{c_{1}}}{2}\frac{\partial }{\partial r}\pm \frac{1}{2}%
\right]   \label{16} \\ J_{0}^{\infty } &=&-i\frac{\partial
}{\partial \phi }.  \notag
\end{eqnarray}
The operators given in (16) close in an $so(2,1)$ algebra. Their
Casimir operator is
\begin{equation}
Q^{\infty }=\frac{1}{4}\left[ c_{1}\frac{\partial ^{2}}{\partial r^{2}}-1%
\right] ,  \label{17}
\end{equation}
we notice that the asymptotic generators obtained from (4) are $p$
independent as one expects. The example developed in the previous
section correspond to $c_{1}=1$. This results are easily obtained
from (14) and (15) when $r\to\infty$. From (1) we obtain that the
asymptotic behavior of the HNP is given by
\begin{equation}
V_{\infty }(z)=\frac{h_{1}+1}{c_{1}},  \label{k18}
\end{equation}
then it is necessary to choose $h_{1}=-1$, in order to satisfy the
conditions given in [1], this condition is satisfied as seen from
(11).

Let us consider the continuous series of $so(2,1)$ for which the
eigenvalues of the Casimir operator, $q=j(j+1)$ are such that $j$
is given by [9]
\begin{equation}
j=-\frac{1}{2}+i\frac{\lambda }{2},\;\lambda \,\ \ \text{real}
\label{19}
\end{equation}
and the compact generator has eigenvalues: $m=m_{0}\pm \sigma ,\;\sigma =$ {%
integer.}

The asymptotic states [6,10] are given by
\begin{equation}
\left| j,m\right\rangle _{\infty }=A_{m}\exp i(\lambda r+m\phi
)+B_{m}\exp i(-\lambda r+m\phi ),  \label{20}
\end{equation}
where the coefficients or the Jost functions $A_{m}$ and $B_{m}$
has to be evaluated. This can be done if we use the expression for
$J_{+}^{\infty }$ given in (16) and then act on the asymptotic
state (20). The result obtained should be compared with the
general expression for the action of generators of an $so(2,1)$
algebra, namely
\begin{equation}
J_{\pm }\left| j,m\right\rangle =\sqrt{(m\mp j)(m\pm j\pm
1)}\left| j,m\pm 1\right\rangle .  \label{21}
\end{equation}

Recursion relations are obtained for $A_{m}$ and $B_{m}$, the
reflection coefficient, $R_{m}=A_{m}/B_{m}$, is found to be
\begin{equation}
R_{m}=\frac{\Gamma (-i\frac{\lambda }{2}+m+\frac{1}{2})\Gamma (i\frac{%
\lambda }{2}+\frac{1}{2})A_{0}}{\Gamma (i\frac{\lambda }{2}+m+\frac{1}{2}%
)\Gamma (-i\frac{\lambda }{2}+\frac{1}{2})B_{0}}.  \label{22}
\end{equation}

Assuming that $A_{0}$ and $B_{0}$ are holomorphic functions of
$\lambda $, it is then easy to verify that the poles of (22)
indeed reproduce the spectra given in\ (12), after using (13). The
general case for the algebraic treatment of the HNP is the same,
since it is  obtained by scaling the variable $r$ as is seen from
(16) and (17). Concerning to an Euclidean connection analysis for
the scattering of the HNP using an $so(2,1)$ algebra see [11].

As a final comment concerning to the scattering sector, the
deformed scattering [12], can be done using the method developed
here and in [11] in simple way, work is in progress.

\section{Satellite potentials}

Let us see an interesting feature concerning the algebraic
description of the bound state sector mentioned before. Let us
denote by $H_{p_{\nu }q_{\nu }}$ the carrier space of
$so(2,1)_{p_{\nu }q_{\nu }}$. Thus the eigenfunctions of HNP
belong to$\ $the direct sum of this spaces with $\nu =0 $ to $\nu
=\nu _{max}$. For a specific state, $\Psi _{p_{\nu }q_{\nu }m_{\nu
}}$ which belongs to a carrier space label by $p_{\nu }$and
$q_{\nu }$, one can ask for the result of the ladder operators
given in (4) acting on this state, the result is [13]:
\begin{eqnarray}
J_{-}\Psi _{p_{\nu }q_{\nu }m_{\nu }} &=&-\frac{\nu (\alpha _{\nu
}-\nu -1-\beta _{\nu })}{1+\beta _{\nu }}\Psi _{_{p_{\nu }q_{\nu
}m_{\nu }}-1} \label{23} \\ J_{+}\Psi _{p_{\nu }q_{\nu }m_{\nu }}
&=&-\beta _{\nu }\Psi _{p_{\nu }q_{\nu }m_{\nu }+1}  \notag
\end{eqnarray}

We obtain states corresponding to different group parameters, this
states are eigenfunctions of a different HNP, we call them
satellite potentials. The problem then is to find their Natanzon
parameters. From (23) we see that
the new state has the same $z$, thus $\left\{ a,\ c_{0},\ c_{1}\right\} $, $%
p_{\nu }$ and $q_{\nu }$ are unchanged while $m_{\nu }\rightarrow
m_{\nu }\pm 1.$ To be more specific, let us deal with the action
of $J_{+}.$ From (8) we obtain
\begin{equation}
\alpha _{s\nu +1}=\alpha _{\nu }+1,\ \beta _{s\nu +1}=\beta _{\nu
}-1,\delta _{s\nu +1}=\delta _{\nu }.
\end{equation}
the index $s$ is used to label the satellite parameters.
Complicated relations for the new set of Natanzon parameters
$f_{s}$, $h_{0s}$, $h_{1s}$ are obtained. For simplicity we study
the potential given in (10) and we add a constant for convenience,
namely
\begin{eqnarray}
V_{PT}&=&-A(A+1){{\mathrm{sech}}}(r)^{2}+B(B-1){\mathrm{csch}}(r)^{2}\nonumber\\
&&+(A-B)^{2}
\end{eqnarray}
In this case the Natanzon parameters are the same ones given in
(11) except that now $h_{1}$ is given by $h_{1}=(-A+B-1)(-A+B+1)$.
The group parameters are the same as in (13). For the energy
spectra it is found
\begin{equation}
E_{PT}(\nu )=-4\nu (\nu -A+B)
\end{equation}

From (8) we obtain
\begin{equation}
\alpha _{\nu }=\frac{2A+1}{2},\ \beta _{\nu }=\frac{2B-1}{2},\
\delta _{\nu }=\frac{A-B-2\nu }{2}
\end{equation}

Let us define $A_{s}$ and $B_{s}$ as the parameters of the
satellite potential, then from (27) and (24) we have
\begin{equation}
A_{s}=A+1,\ B_{s}=B-1
\end{equation}

The energy spectra of the satellite potential is obtained from \
the last equation in (8) and the result is
\begin{equation}
E_{s}(\nu +1)=E_{PT}(\nu )-h_{s1}-(A-B)^{2}
\end{equation}

We notice that the change of parameters obtained in (28) are not
the same occurring in SUSYQM [14], neither in the $so(2,2)$
potential approach [6]. To see clearly this effect, consider the
case where $B=0.$ Thus, with this technique one generates HNP
potentials \ from a seed that can be analyzed by the algebraic
technique developed in [4]. The other cases of shape invariant
potentials can easily done by the same method developed in this
note.

\vspace*{1cm}
\noindent{\large\bf{Acknowledgments}}

This work has been partially supported by grants USB-61D30 
and FONACIT No. 6-2001000712.


\begin{thebibliography}{99}
\bibitem{k1}  Natanzon G. A., \textit{Theor. Mat. Fiz.}, \textbf{38} (1979)
146; Vestn. Leningr. Univ., \textbf{6} (1977) 33. English
traslation: physics/9907032

\bibitem{k2}  Wu J., Alhassid Y. and G\"{u}rsey F., \textit{Ann. Phys.}
\textbf{196} (1989) 163.

\bibitem{k3}  Cordero P. and Salam\'{o} S., \textit{Found. Phys.} \textbf{23
}(1993) 675; \textit{J. Math. Phys.} \textbf{35} (1994) 3301

\bibitem{k4}  Cordero P. and Salam\'{o} S., \textit{J. Phys. A: Math.Gen.}
\textbf{24} (1991) 5299.

\bibitem{k5}  Zwanziger, D.\textit{\textrm{, J. Math. Phys., }}8 (1967)
1858; Barut A. O. and Rasmussen, W., \textit{J. Phys. B: Atom.
Molec.}
\textbf{6 }(1973) 1965; Barut A. O., Rasmussen, W. and Salam\'{o} S., \textit{%
Phys. Rev. D}, \textbf{10 }(1974) 622; \textit{Phys. Rev. D},
\textbf{10} (1974) 630; Rasmussen W. and Salam\'{o}
S\textit{\textrm{., J. Math. Phys., \textbf{2}}}0 (1979) 1064.

\bibitem{k6}  Alhassid Y., G\"{u}rsey F. and Iachello F, \textit{Ann. Phys.}
\textbf{167} (1986) 181; Wu J., Iachello F and Alhassid Y.,
\textit{Ann. Phys.} \textbf{173 }(1987) 68.

\bibitem{k7}  Kerimov, G. A., \textit{Phys. Rev. Lett., }\textbf{90} (1998)
2976; Kerimov, G. A. and Sezgin M. \textit{J. Phys. A: Math. Gen. }\textbf{31%
} (1998) 7901.

\bibitem{k8}  Frank A. and Wolf K. B., \textit{Phys. Rev. Lett.} \textbf{32 }%
(1984) 1737.

\bibitem{k9}  Barut, A. O. and Raczka \textit{Theory of Group
Representations and Applications,\textrm{\ }}PWN\textrm{, Warszawa
1977.}

\bibitem{k10}  Taylor, J. R\textit{\textrm{., Scattering Theory: The Quantum
Theory on Nonrelativistic collisions, }}John Wiley \& Sons., 1972.

\bibitem{k11}  Albrecht H. and Salam\'{o} S., accepted for publication in 
Rev. Mex. F\'{\i}sica, 2002, proceedings of ``III
Congreso de la Sociedad Venezolana de F\'{i}sica'', 10-14 diciembre 2001,
Caracas, Venezuela.

\bibitem{k12}  Frank A., Alonso C. E. and G\'{o}mez-Camacho J., Rev. Mex. de
F\'{\i}sica \textbf{39} (1993), Sup. 2, 64.

\bibitem{k13}  Codriansky S. and Salam\'{o} S., {\it{J. Phys. A: Math. Gen.}} 
{\bf{35}},19, 4269 (2002), quant-ph/0109085; \textit{J.
Phys. A: Math. Gen. }\textbf{32} (1999) 6287.

\bibitem{k14}  Cooper F., Ginocchio J. N. and Khare A., \textit{Phys. Rev. D}
\textbf{36} (1987) 2458
\end{thebibliography}
\end{document}